\documentclass[aps,preprint]{revtex4}

\usepackage{amsmath}
\usepackage{bm}
\usepackage{graphicx}
\usepackage{hyperref}
\newcommand{\vvr}{\mathbf{r}}

\newcommand{\vcr}{\mathbf{R}}

\newcommand{\brho}{\bm\rho}

\newcommand{\vra}{\mathbf{R}_{12}}
\newcommand{\vrb}{\mathbf{R}_{23}}
\newcommand{\vrc}{\mathbf{R}_{31}}
\newcommand{\req}{R^{(e)}}

\graphicspath{{picqd/}}

\begin{document}

\title[On the symmetry in the three-electron quantum dot]%
{On the symmetry of the electronic density in the 
three-electron parabolic quantum dot}
\author{A.V.~Meremianin}
\date{\today}
\affiliation{General Physics Department, Voronezh State University,
394006, Voronezh, Russia}

\begin{abstract}
The structure of the lowest states of a three-electron axially
  symmetric parabolic quantum dot in a zero magnetic field is investigated.
It is shown that the electronic density of the quartet ${}^{4}S$-states
possesses 
certain approximate symmetry which is best seen when using Dalitz plots as the
visualization tool.
It is demonstrated that the origin of that symmetry is caused by the symmetry
of the potential energy in the vicinity of its minimum.
The discovered symmetry could provide an insight into the problem of the
separation of slow and fast variables in the Schr\"odinger equation for the
axially or spherically symmetric quantum dots.
\end{abstract}

\maketitle

\section{Introduction}
\label{sec:intro}

Symmetries play very important role in physics
\cite{fano96:_symmet_quant_phys}.
For example, such fundamental physical principles as the energy, linear and
angular momentum conservation laws are based on the symmetry of the space-time
with respect to translations and rotations.
These symmetries can be considered as ``kinematical symmetries'' rather that
dynamical ones since the former are independent on the interparticle
interaction.
It turns out that every exactly solvable quantum-mechanical problem possesses
some kind of dynamical symmetry.
Apparently, such situations are rather rare and are well-studied.
Examples include the harmonic oscillator
\cite{fock28:_fock_darwin,PhysRev.57.641,0305-4470-11-2-005}
and hydrogen atom problems \cite{fock35:_o4,landau3tom_eng}.

In some cases, however, a quantum system may have approximate symmetries.
The existence of approximate symmetries leads to significant simplification
of the analysis of the problem.
Approximate symmetries are hard to find since the operator of the
corresponding symmetry transformation does not commute with the Hamiltonian
\cite{ushveridze2017quasi}.
Nevertheless, approximate symmetries have been found for such nonseparable
problems as hydrogen atom in a uniform magnetic field
\cite{PhysRevLett.45.1092} and doubly excited states of helium atom
\cite{0953-4075-24-20-004}.
Recently, the approximate symmetries of the nodal lines of the lowest 
\cite{bressanini05:_jcp_nodes,bressanini05:_nodes_he} and resonant
doubly-excited states \cite{0953-4075-24-10-004} of the helium atom have been
discovered. 

In the present article the approximate symmetry of the wave functions of the
three electrons subjected to the circularly symmetric parabolic potential
possesses is found.
This problem is of interest since it is relevant in the theoretical
investigation of the electronic structure of quantum dots.
These are semiconductor structures which can confine electrons 
\cite{0034-4885-64-6-201}.
Therefore, they are often referred to as ``artificial atoms''
\cite{RevModPhys.74.1283}.
The theoretical study of few-electron quantum dots allows
one to analyze the role of electronic correlations in nanostructures
\cite{PhysRevLett.79.3475,macsym00:_eckart_frame_dots,puente:125315}.

The electronic structure of quantum dots can often be described by the model
in which electrons having an effective mass move in a parabolic confining
potential \cite{Fang2007551,0253-6102-48-6-030,PhysRevB.65.115312}.
% Often, a quantum dot can be modeled as a two-dimensional system of electrons
% having effective mass and confined via the axially symmetric parabolic
% potential.
Quantum dots with one or two-electrons are comparatively simple to study since
the corresponding theory can be developed using various analytical model
approaches 
\cite{puente:125315,rashid03:_2_electron_quant_dots}.
Theoretical investigations of many-electron quantum dots are much more
complicated because they require solution of many-dimensional partial
differential equations which cannot be done analytically.
Few-electron quantum dots are particularly difficult to study as
in this case the application of various mean field approximations cannot be
justified.

Obviously, the three-electron parabolic quantum dots are the simplest
few-electron objects to analyze.
They were studied rather extensively during the last decades
\cite{PhysRevLett.82.3320,PhysRevB.65.115312,
simonovic06:_three_electr_2d_qdot}.
In particular, much attention has been paid to the properties of the
three-electron quantum dots in a magnetic field
\cite{0953-8984-21-7-075302,Fang2007551}.
In the mentioned papers the energy spectrum was calculated using various
approaches and the structure of the electronic density was studied using
pair-correlation functions.
The latter, however, is not always appropriate as it could hide some
interesting features of electronic density which are caused by triple
correlations.
In the three-body problem it is more instructive to analyze the structure of
the electronic density directly, using some suitable set of internal
variables.
The treatment of the present article is based on the Dalitz-plot technique
which is often used to analyze the angular distributions
in three-body break-up processes in particle and molecular physics
\cite{dalitz53:_plots,helm99:_h3_first,galster_helm05:_h3_results,%
nature08:_dots_mol_states}.

The use of Dalitz plots for the visualization of the electronic density
greatly simplifies the analysis of its symmetries.
% of the wave function of the circularly symmetric three-electron parabolic
% quantum dot.
Below it is shown that the Dalitz plots corresponding to the ground (and
lowest excited) quartet states of the three-electron parabolic quantum dots
posses some approximate symmetry similar to that observed in the model of the
break-up of a three-body rigid rotator \cite{meremianin06:_h3_kinem_model}.
This symmetry means that, at a given value of the hyperradius which defines
the overall ``size'' of the configuration triangle, the dependence of the
electronic density on the area of that triangle is very much stronger than on
its shape.
The detailed analysis performed with the help of internal variables similar to
``Dalitz-Fabri'' coordinates
\cite{0022-3700-15-23-001,krivec98:_few_bod_hypsph} explains the origin of
the observed approximate symmetry (Sec.~\ref{sec:sym-origin}).
Namely, it is caused by the symmetry of the total potential in the vicinity of
its local minimum.
This symmetry can be uncovered by taking the power series
expansion of the potential.
For some particular values of the confinement strength the magnitude of the
distortion of the symmetry is estimated in Sec.~\ref{sec:numerical-results}.

For the sake of brevity, numerical calculations were carried out only for 
states with zero orbital momentum including the ground states of
the quantum dot in the absence of external fields.

%% It is known that the wave function 

\section{The Hamiltonian of the three-electron quantum dot}
\label{sec:hamilt-three-electr}

The Schr\"odinger equation for the three electrons moving in a two-dimensional
parabolic quantum dot is
\begin{eqnarray}
  \label{eq:ge-1}
&& -\frac{\hbar^2}{2m_e} 
\left( \sum_{i=1}^3 \Delta_{\vcr_i} + U
\right)\, \Psi_t = E\, \Psi_t, \\
%%%%%%%%%%%%%
&& U = \sum_{i=1}^3 \frac{m_e\, \omega^2\, R^2_i }{2}
 + \sum_{i>j=1}^3 \frac{e^2}{\epsilon\, |\vcr_i - \vcr_j|},
\end{eqnarray}
where $m_e$ is the effective electron mass and
$\epsilon$ is the dielectric constant.

%% \textit{We forget about the transversal magnetic field for a while.}

\begin{figure}[ptbh]
\centering
 \includegraphics[width=6cm]{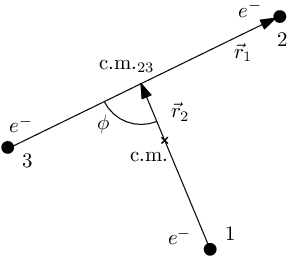}
 \caption{Jacobi vectors for the three-body system. $\mathrm{c.m.}_{23}$
   is the c.m. of the electrons ${2}$ and ${3}$.}%
\label{fig:3bd-jacobi}%
\end{figure}

For the confinement potential given in (\ref{eq:ge-1}) it is possible to
separate out the motion of the c.m. of electrons by introducing two Jacobi 
vectors $\vvr_{1,2}$ as is shown in Fig.~(\ref{fig:3bd-jacobi}).

The kinetic energy operator in terms of Jacobi vectors can be written as
\begin{equation}
  \label{eq:kin-op}
- \frac{\hbar^2}{2 m_e} \sum_{i=1}^3 \frac{\partial^2}{\partial \vcr_i^2}
= -\frac{\hbar^2}{6 m_e} \frac{\partial^2}{\partial \vcr_{c.m.}^2}
- \frac{\hbar^2 }{m_e} \frac{\partial^2}{\partial \vvr_1^2}
- \frac{3\,\hbar^2 }{4 m_e} \frac{\partial^2}{\partial \vvr_2^2}.
\end{equation}
The sum of squared lengths of the position vectors re-written via Jacobi
vectors is diagonal,
\begin{equation}
  \label{eq:sqred-sum}
R^{2} \equiv R_1^2+R_2^2+R_3^2 = 3 R_{c.m.}^2 
           + \frac{1}{2} r_1^2 + \frac{2}{3} r_2^2.
\end{equation}

It is convenient to introduce the mass-scaled Jacobi vectors
by making the following replacements in (\ref{eq:kin-op}) and
(\ref{eq:sqred-sum}):
\begin{equation}
  \label{eq:repl-1}
 \vvr_1 \to \sqrt 2\, \vvr_1, \quad
 \vvr_2 \to \sqrt\frac{3}{2} \, \vvr_2.
\end{equation}
With these replacements the Schr\"odinger equation reads
\begin{equation}
  \label{eq:se-2}
(H_{c.m.} + H_{int})\, \Psi_{t} = E \, \Psi_{t},
\end{equation}
where $H_{c.m.}$ is the Hamiltonian describing the motion of c.m. of three
electrons,
\begin{equation}
  \label{eq:H_cm}
   H_{c.m.} = -\frac{\hbar^2}{6 m_e} \frac{\partial^2}{\partial \vcr_c^2}
      + \frac{3 m_e\, \omega^2\, R_{c.m.}^2}{2},
\end{equation}
and $H_{int}$ is the Hamiltonian corresponding to the internal (relative)
motion of electrons in the parabolic trap
\begin{equation}
  \label{eq:H_3e}
  H_{int} = -\frac{\hbar^2}{2 m_e} \left( \Delta_1 + \Delta_2 \right)
  + \frac{m_e\, \omega^2\, (r_1^2+r_2^2)}{2} 
  + e^2 \kappa\, U_{cl},
\end{equation}
where $\Delta_{1,2} = \partial^2/\partial \vvr_{1,2}^2$,
$\kappa=1/\epsilon$ and $U_{cl}$ denotes the Coulomb repulsion terms
\begin{equation}
  \label{eq:def-u}
  U_{cl} =  \frac{1}{\sqrt 2 \, r_1}
+ \frac{\sqrt 2}{|\vvr_1 + \vvr_2 \sqrt{3}|}
+ \frac{\sqrt 2}{|\vvr_1 - \vvr_2 \sqrt{3}|}.
\end{equation}

%% By replacing 
Dividing the Schr\"odinger equation by $\hbar \omega$ it can be brought to
dimensionless form by making the replacements
$r_{1,2} \to r_{1,2} \sqrt{\hbar/(m_e \omega) } $.
As a result, the Hamiltonian assumes the form
\begin{equation}
  \label{eq:h-1}
  H_{int} = -\frac{\Delta_1 + \Delta_2}{2}   
+ \frac{r_1^2+r_2^2}{2} + R_c \, U_{cl},
\end{equation}
where the variables $r_{1,2}$ are dimensionless and $R_c$ is the Coulomb
strength parameter,
\begin{equation}
  \label{eq:rc}
  R_c = \frac{e^{2} \kappa}{\hbar} \sqrt\frac{m_e}{\hbar \omega}
= \alpha \kappa \sqrt{\frac{m_e c^{2}}{\hbar \omega}},
\end{equation}
where $\alpha$ is the fine structure constant.
The numerical calculations were carried out with the effective electron mass
$m_e = 0.067 m$ and $\kappa=12.4$, which correspond to GaAs, so that 
\begin{equation}
  \label{eq:r-2}
  R_c = \frac{3.443}{\sqrt{(\hbar \omega)_\mathrm{mEv}}}.
\end{equation}

%%%%%%%%%%%%%%%%%

\section{Dalitz plots of the electronic density}
\label{sec:dalitz-plots-density}

According to (\ref{eq:se-2}) the total wave function $\Psi_t$ can be expressed
as the product of the wave function $\Psi_{c.m}$ describing the motion of 
c.m. of three electrons and the wave function $\Psi$ describing the
relative motion of electrons:
\begin{equation}
  \label{eq:psi-prod}
  \Psi_t = \Psi_{c.m.} (\vcr_{c.m})\, \Psi (\vvr_1,\vvr_2).
\end{equation}
The wave function $\Psi_{c.m.}$ has the same form as the wave function of a
harmonic oscillator with the mass $3 m_e$.
It is the internal wave function which is determined by the electronic
correlations.
Therefore, below only the electronic density $D= | \Psi
(\vvr_1,\vvr_2)|^{2}$ is considered.

In the three-electron quantum dot the density $D$ depends on three
internal variables $\xi=(\xi_1, \xi_2,\xi_3)$. 
Thus, $D(\xi)$ is a surface in the four-dimensional space and as such
cannot be visualized.
However, if we fix one of the internal variables, say $\xi_1$, then the
function $D(\xi_1=const, \xi_2, \xi_3)$ becomes a three-dimensional
surface which can be depicted on a sheet of paper as a color intensity map.
Since the hyperradius $R$ is independent of the particle exchange, it is 
convenient to visualize $D (\xi)$ as a series of 3d surfaces with variable
values of $R=0, \ldots, R_{max}$.
Now the question is how to choose the two remaining internal variables to
facilitate the features of the electronic density.
Below we will use two dimensionless internal coordinates similar to those
of a Dalitz plot.

Conventional Dalitz plots are the diagrams which depict the angular
distributions of linear momenta of three particles
\cite{dalitz53:_plots,helm99:_h3_first}. 
Originally, they were introduced to visualize the angular distributions of
$K$ mesons in particle physics \cite{dalitz53:_plots}.
On the Dalitz plot, each configuration of particle's
momenta is represented by the point inside a circle so that the exchange of
particles is equivalent to the rotation by the angle $(2/3)\pi$ with respect
to the center of the plot which itself corresponds to the equilateral
configuration when vectors of particle's momenta form an equal-side triangle.
Points on the edge of the circle describe collinear configurations
when particles fly apart along the same line.

To apply the Dalitz plot technique to the analysis of the electronic density 
we choose the coordinates of the polar plot to be the Dalitz coordinates
\cite{dalitz53:_plots,springerlink:10.1007/BF02781042}
in the two-dimensional configuration space
\begin{equation}
  \label{eq:dp-xy}
x = \frac{R_1^{2} - R_2^{2}}{\sqrt3\, R^{2}}, \quad
y=\frac{1}{3} - \frac{R^{2}_3}{R^{2}}.
\end{equation}
Here, it is assumed that c.m. of the three electrons is located at the origin
of the coordinate frame, i.e.
\begin{equation}
  \label{eq:dp-1}
%%  R_1^{2} + R_2^{2} + R_3^{2} = R^{2},
\vcr_{c.m.} = \vcr_1 + \vcr_2 + \vcr_3=0.
\end{equation}
The Dalitz coordinates (\ref{eq:dp-xy}) can also be expressed in terms of
mass-scaled Jacobi vectors:
\begin{equation}
  \label{eq:dp-4a}
  \begin{split}
    x &= \frac{1}{2 \sqrt 3 \, R^{2}} \left( r_2^{2} - r_1^{2}
      - \frac{1}{\sqrt 3} (\vvr_1  \cdot \vvr_2) \right), \\
%%%%%%%%
    y &= \frac{1}{6 \, R^{2}} \left( r_2^{2} - r_1^{2} 
        + 2 {\sqrt 3} (\vvr_1 \cdot \vvr_2) \right),
  \end{split}
\end{equation}
where the hyperradius $R = r_1^{2} + r_2^{2}$.

In literature are often used the symmetry adapted hyperspherical
coordinates also known as ``Dalitz-Fabri coordinates'', $R,a,\lambda$,
  \cite{PhysRev.51.655,0022-3700-15-23-001,krivec98:_few_bod_hypsph,%
badalyan1966three} which are defined by the equations
\begin{equation}
  \label{eq:a1}
  \begin{split}
    r_2^{2} - r_1^{2} &= R^{2} \sin a \, \cos \lambda, \\
    (\vvr_1 \cdot \vvr_2) &= \frac{R^{2}}{2} \sin a \, \sin \lambda,
  \end{split}
\end{equation}
where $0 \le a \le \pi/2$ and $0 \le \lambda \le 2 \pi$.

Note that the hyperangles $a,\lambda$ were, in fact, originally introduced by
Gronwall and published in his posthumous paper \cite{PhysRev.51.655} where the
Hamiltonian of the helium atom \cite{PhysRev.51.655} was written in terms of
the variables $R,a,\lambda$.
Therefore, below these coordinates will be referred to as
``Gronwall-Dalitz-Fabri coordinates'' (GDF).

Coordinates having similar kinematical properties as GDF
coordinates $a,\lambda$, %% (\ref{eq:a1}) 
were used in molecular physics by several authors
including  Kuppermann \cite{Kuppermann1975374}, Mead
\cite{mead92rmp:berry_phase}, Pack \cite{pack91:_3bd_coord}.
Mishra and Linderberg \cite{Linderb_h3_pes_dalitz} used Mead coordinates to
visualize potential energy surfaces in triatomic molecules.

From (\ref{eq:dp-4a}), (\ref{eq:a1}) one can deduce the connection of
Cartesian coordinates $x,y$ to hyperangles $a, \lambda$,
\begin{equation}
  \label{eq:xy-al}
  \begin{split}
    x &%
    % = \frac{1}{3} \sin \alpha \, \left(
    %   \frac{\sqrt 3}{2} \cos \lambda
    %   - \frac{1}{2} \sin \lambda \right)
    = \frac{\sin \alpha }{3}
    \, \cos \left(\lambda+ \frac{\pi}{6} \right), \\
%%%%%%%%
    y &= \frac{\sin \alpha }{3} \, \sin \left(\lambda + \frac{\pi}{6} \right),
  \end{split}
\end{equation}
From these equations it follows that the polar radius $\rho$ on the Dalitz
plot is
\begin{equation}
  \label{eq:rho-xy}
  \rho \equiv \sqrt{x^{2}+y^{2}} = \frac{\sin a}{3}.
\end{equation}
Using the identities (\ref{eq:sqred-sum}), (\ref{eq:dp-xy}), (\ref{eq:dp-1})
we obtain the representation of $\rho$ in terms of position vectors:
\begin{equation}
  \label{eq:dp-2}
\rho^{2} = \frac{1}{9} - \frac{4\, | \vcr_1 \times \vcr_2|^{2}}{3 R^{4}}.
\end{equation}

From (\ref{eq:dp-2}) and (\ref{eq:dp-1}) it is seen that the polar radius
$\rho$ of the Dalitz plot is invariant under the particle exchange.
This means that the exchange of particles is equivalent to the rotation or
reflection of the diagram.

Expression (\ref{eq:dp-2}) written in terms of Jacobi vectors has the form
\begin{equation}
  \label{eq:dp-2c}
\rho^{2} = \frac{1}{9} 
- \frac{4\, | \vvr_1 \times \vvr_2|^{2}}{27 R^{4}}.
\end{equation}
The geometrical meaning of this equation is that polar radius of
the Dalitz plot is determined by the ratio of the area $S$ of the
configuration triangle and the hyperradius R:
\begin{equation}
  \label{eq:dp-3a}
  \rho^{2} = \frac{1}{9} - \frac{16\,S^{2}}{27 R^{4}}. 
\end{equation}
Indeed, the positions of particles define the vertices of the configuration
triangle whose area is
\begin{equation}
  \label{eq:dp-3}
  S = \frac{1}{2} | \vvr_1 \times \vvr_2 |.
\end{equation}

%%%%%%%%%%%%%%%%%%%%%%%%%%

\section{The numerical results}
\label{sec:numerical-results}

The wave function of the three electrons was obtained by diagonalizing the
Hamiltonian (\ref{eq:h-1}) in the basis of Fock-Darwin states 
\cite{darwin30:_fock_darwin,fock28:_fock_darwin} 
which are defined by
\begin{eqnarray}
  \label{eq:fd1}
  \Psi_{n,m} (\vvr) &=& \frac{e^{-im \phi}}{\sqrt{2\pi}}\, \psi_{n,m}(r), \\
\psi_{n,m}(r) &=& %% \frac{1}{\lambda} 
\sqrt\frac{n!}{(n+|m|)!}
\left(
\frac{r}{\sqrt2 } \right)^{|m|} %% \nonumber \\
%% & \times &
 e^{-r^2/4} L^{|m|}_n \left( \frac{r^2}{2 } \right),
\end{eqnarray}
where $L^{|m|}_n$ is the associated Laguerre polynomial \cite{Bateman-II}.
Fock-Darwin wave functions (\ref{eq:fd1}) diagonalize the Hamiltonian of an
electron in a parabolic circular trap.
%% The parameter $\lambda^2=\hbar/(2m_e \omega)$, where 
%% $\Omega=\sqrt{\omega_0^2 + \omega_c^2/4}$, and $\omega_c=eB/m^*$ is the
%% cyclotron frequency.
The corresponding single-electron energy is (in units of $\hbar \omega$)
\begin{equation}
  \label{eq:fd2}
  E_{n,m} = (2n+1+|m|). %% \hbar \omega.
%%  E_{n,m} = (2n+1+|m|) \hbar \Omega - m \hbar \omega_c/2.
\end{equation}

The expansion of the wave function of an $S$-state over the Fock-Darwin
states has the form
\begin{equation}
  \label{eq:e-1a}
 \Psi (\vvr_1, \vvr_2) =
\sum_{n=0}^{N}  \sum_{n'=0}^{N} \sum_{m=-m_0}^{m_0}
F_{n n', m}\, \Psi_{n,m}(\vvr_1)\, \Psi_{n', -m} (\vvr_2),
\end{equation}
where $N$ and $m_0$ determine the accuracy of the representation of the wave
function.
The number of terms in the expansion (\ref{eq:e-1a}) is 
\begin{equation}
  \label{eq:e-2a}
  Z_{N m_0} = (N+1)^2\, (2 m_0+1).
\end{equation}
The satisfactory convergence was achieved at $N,m_0 \sim 5-7$. 
(Note that large values of $N,m_0$ lead to occurrence of spurious
oscillations which degrades the accuracy of computations
\cite{haftel1989fast}.)
Obtained results for the energy of the ground states are in good agreement
with existing in literature \cite{0253-6102-48-6-030}.

\begin{figure}
  \centering
 \includegraphics[width=8cm]{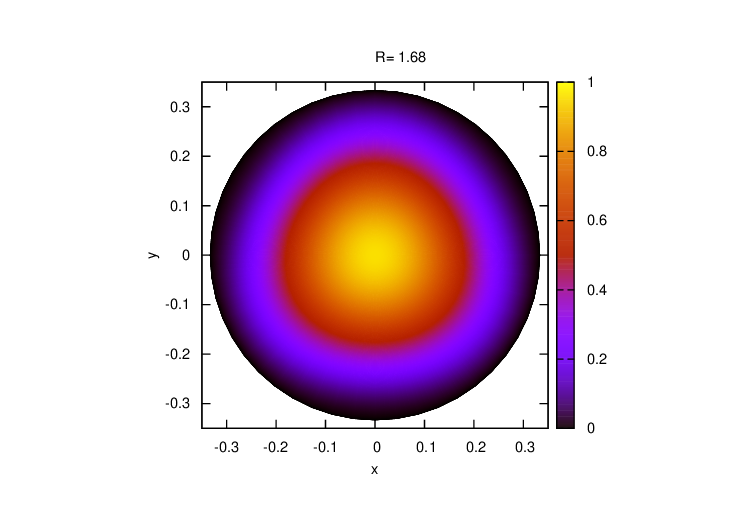}
 \includegraphics[width=8cm]{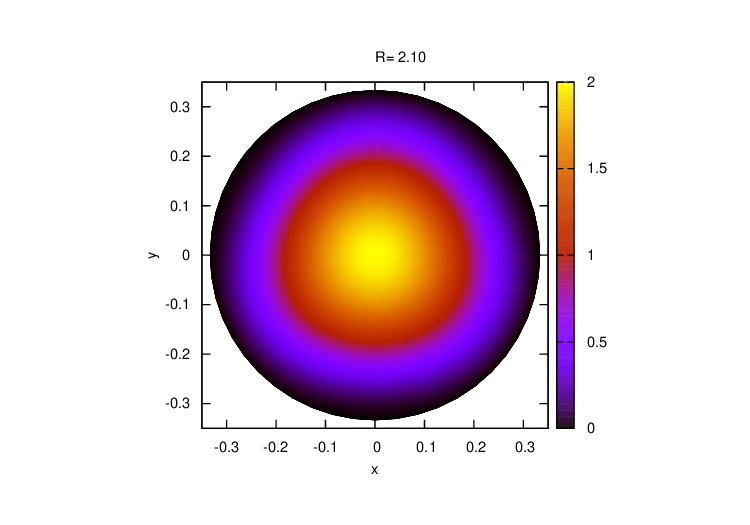}
  \caption{(Color online) The Dalitz plot for the electronic density of the
    ground state at the confinement 
    energy $\hbar \omega = 0.40$ meV ($R_c=5.444$) for two values of
    hyperradius $R$.}
  \label{fig:dp1}
\end{figure}

Dalitz plots of the ground state electronic density 
$D_{0} = | \Psi_{0} (R, a, \lambda) |^{2}$ 
are given in Fig.~\ref{fig:dp1} for the Coulomb strength parameter
$R_c=5.444$ (which corresponds to the confinement energy $\hbar \omega = 0.40$
meV) and two values of the hyperradius $R$.
As is seen, the density has maximum at the center  of the plot which is the
equilateral configuration and decreases as the configuration triangle becomes
more prolate, finally vanishing for collinear configurations.

\begin{figure}
  \centering
  \includegraphics[width=8cm]{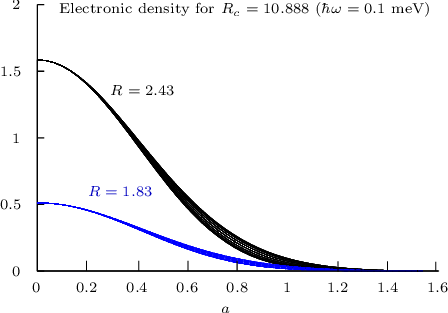}
  \includegraphics[width=8cm]{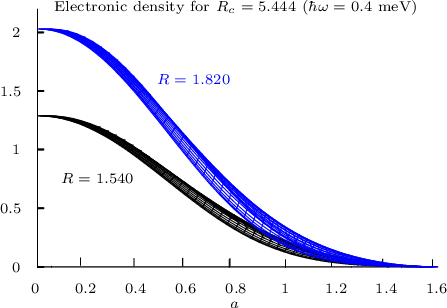}
  \includegraphics[width=8cm]{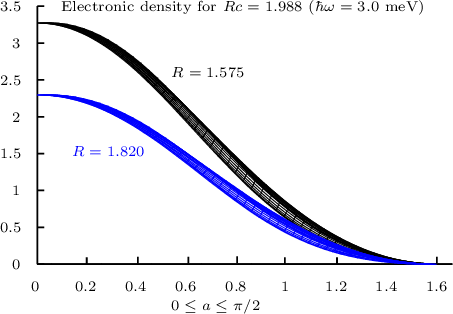}
  \caption{(Color online) Projections of the electronic density 
     $D_{0}=|\Psi_{0} (R, a, \lambda)|^{2}$ of the quartet $S$-state on
     the plane $\lambda=const$ for three values of the Coulomb strength
     parameter $R_c$.
   Note that in the case of independence of the density on the hyperangle
   $\lambda$ the projections would be single thin lines.}
  \label{fig:albd-1}
\end{figure}

The striking feature of the diagrams in Fig.~\ref{fig:dp1} is the remarkably
weak dependence of the density on the polar angle of the plot.
In order to estimate the magnitude of this dependence, Fig.~\ref{fig:albd-1}
shows the projection of the density $D_{0} (R, a, \lambda)$ on the surface
$\lambda=const$ of the Cartesian frame with coordinates 
$(x_1,x_2,x_3)=(a, \lambda, D_0)$. 
The width of the curves shown in Fig.~\ref{fig:albd-1} is 
determined by the variation of the density as a function of the polar
angle (which is actually $\lambda+\pi/6$, see (\ref{eq:xy-al})).
The structure of the electronic density shown in
Figs.~\ref{fig:dp1},\ref{fig:albd-1} is preserved also for other values of the
hyperradius $R$.
%% , \cite{video1}, \cite{video2}.
For larger values of the confinement energy $\hbar \omega$ (which
means smaller Coulomb parameter $R_c$) the electronic
density has more pronounced maxima at the equilateral configuration.
The calculations were also performed for other values of the confinement
energy in the range $0.1$ -- $0.4$ meV. 
In all cases the symmetry of the electronic density of the quartet states is
essentially the same as in Figs.~\ref{fig:dp1},\ref{fig:albd-1}.

Note that at the values of the confinement energy $\hbar \omega \ge 0.62$ meV
the ground state of the three-electron quantum dot is the doublet
${}^{2}P$-state with the total spin $S=1/2$ and the total orbital momentum
$L=1$ \cite{PhysRevB.65.115312,0253-6102-48-6-030}.
At $\hbar \omega = 0.62$ meV the transition to the ground quartet
${}^{4}S$-state ($S=3/2$, $L=0$) occurs which is often referred to as the
formation of the 
``Wigner molecule'' 
\cite{PhysRevLett.82.3320,PhysRevB.63.113313,macsym00:_eckart_frame_dots,PhysRevB.65.115312}.

The Dalitz plots corresponding to the excited quartet states also have
circular symmetry similar to that shown in Fig.~\ref{fig:dp1}.
However, the computations of the wave functions for the excited states are
less accurate than those of their eigenvalues and the corresponding results are
not shown here.

%%%%%%%%%%%%%%%%%%%%%%%%%%%

\section{Expansion of the Coulomb energy and the origin of the symmetry}
\label{sec:sym-origin}

In terms of position vectors
the potential energy of the electron-electron interaction in the quantum dot
reads ($e=1$)
\begin{equation}
  \label{eq:uee}
  U = \frac{1}{R_{12}} + \frac{1}{R_{23}} + \frac{1}{R_{31}},
\end{equation}
where $R_{ij} = | \vcr_i - \vcr_j|$.

According to Earnshaw's theorem, the potential energy of the system of
particles interacting via Coulomb forces cannot have minimum.
However, in the case of electrons in a parabolic trap the equilibrium
configurations can exist, i.e. there are minima of the \textit{total}
potential energy.
Thus, we can expand the potential (\ref{eq:uee}) in the vicinity of
the equilibrium configuration.
For example, the power series expansion of the first term on rhs of
(\ref{eq:uee}) can be written as 
\begin{multline}
  \label{eq:ra1}
  \frac{1}{R_{12}} = \sum_{k=0}^{\infty} \frac{1}{k!} 
( (\vra-\vra^{(e)}) \cdot \nabla)^{k} \, \frac{1}{r} \,
\biggr|_{r=\req_{12}} = \\
= \frac{1}{\req_{12}}
+ \left(
\frac{1}{\req_{12}}
- \frac{ (\vra \cdot \vra^{(e)})}{{\req_{12}}^{3}}
\right) \\
%%%%%%%%%%%%%%%
+ \left(
 \frac{1}{\req_{12}} 
- \frac{R_{12}^{2} + 4 (\vra\cdot \vra^{(e)})}{2{\req_{12}}^{3}}
+ 3 \frac{(\vra\cdot \vra^{(e)})^{2}}{2{\req_{12}}^{5}}
\right)
+ \ldots,
\end{multline}
where $\vcr^{(e)}_{ij}$ is the position vector pointing from $i$-th to
$j$-th electron at the equilibrium configuration. 
For the equilibrium configuration being an equilateral triangle we have that 
$\req_{12} = \req_{23} = \req_{31} \equiv R_e$.
If we take into account only zero- and first-order terms in the
expansion (\ref{eq:ra1}) then the Coulomb potential (\ref{eq:uee}) becomes
\begin{equation}
  \label{eq:uee-r}
  U = \frac{6 }{R_e} 
- \frac{ (\vra \cdot \vra^{(e)}) + (\vrb \cdot \vrb^{(e)})
+ (\vrc \cdot \vrc^{(e)})}{R_e^{3}}.
\end{equation}
We have to specify also the mutual orientation of the two configuration
triangles, one built on equilibrium mutual vectors 
$\vra^{(e)}, \vrb^{(e)}, \vrc^{(e)}$ and
another one built on the instantaneous vectors $\vra, \vrb, \vrc$.
This can be done by using the moving frame which satisfies the Eckart
condition \cite{eckart35:_eckart_frame},
\begin{equation}
  \label{eq:eckart2c}
  [\vra^{(e)} \times \vra] + [\vrb^{(e)} \times \vrb]
 + [\vrc^{(e)} \times \vrc]  =0.
\end{equation}
In terms of mass-scaled Jacobi vectors this equation reads
\begin{equation}
  \label{eq:eckart2c-ms}
  [\brho_1 \times \vvr_1] + [\brho_2 \times \vvr_2] =0.
\end{equation}
where $\brho_{1,2}$ are the equilibrium mass-scaled Jacobi vectors.
As a result, the potential energy (\ref{eq:uee-r}) assumes the form
\begin{equation}
  \label{eq:uee-rms}
    U = \frac{6 }{R_e} 
- 3\, \frac{ (\brho_1 \cdot \vvr_1 ) + (\brho_2 \cdot \vvr_2)}{R_e^{3}}.
\end{equation}

In the Eckart frame the sum $(\brho_1 \cdot \vvr_1 ) + (\brho_2 \cdot \vvr_2)$
defines the Eckart parameter ${\cal F}$ which can be written as
\cite{meremianin04:_eckart,avm_jomc13}
\begin{equation}
  \label{eq:F-param}
  {\cal F} = \sqrt{(\rho_1 r_1)^{2} + (\rho_2 r_2)^{2}
+ 2 \rho_1 \rho_2 r_1 r_2 \, \cos (\phi - \phi_e) },
\end{equation}
where $\phi_e$ is the angle between vectors $\brho_1$ and $\brho_2$.
For the equilibrium configuration being an equilateral triangle we
have that $\phi_e =\pi/2$ and $\rho_1 = \rho_2 = R_e/\sqrt 2$ and
the above identity becomes
\begin{equation}
  \label{eq:F-param-a}
  {\cal F} = R_e \sqrt{(r_1^{2} + r_2^{2})/2 + r_1 r_2 \, \sin \phi}.
\end{equation}
Using (\ref{eq:a1}) one can derive the expression for the 
Eckart parameter in terms of GDF variables:
\begin{equation}
  \label{eq:F-param-b}
  {\cal F} = \frac{R_e R}{2} \sqrt{2 + 3 \cos^{2} a}.
\end{equation}
Thus, the potential energy (\ref{eq:uee-rms}) evaluates to
\begin{equation}
  \label{eq:u-3}
  U = \frac{3}{2 R_e} 
\left( 4 -  \frac{R}{R_e} \sqrt{2 + 3 \cos^{2} a} \right).
\end{equation}
As is seen, the potential energy does not depend on the second hyperangle
$\lambda$ and, hence, the dependence of the wave function on $\lambda$ is
caused by the contribution of higher order terms in the expansion of the
Coulomb potential (\ref{eq:ra1}).
The results of numerical calculations presented above allows one to estimate
the contribution of the higher-order multipoles in the expansion of Coulomb terms
to be less than $10\%$ for the chosen values of the electron effective mass
and the confinement strength.

%%%%%%%%%%%%%%%%%%%%%%%%%%%

\section{Conclusion}
\label{sec:conclusion}

In the presented article the symmetry of the electronic density of the
circular parabolic three-electron quantum dots has been investigated.
It is found that the electronic density (and the wave functions) of the
quartet states depends on the shape of the configuration triangle much weaker
than on its overall size and area.
Such property of the density can be understood by employing the power
(i.e. multipole) expansion of the total potential energy around
the equilibrium configuration.
The mentioned symmetry is best seen in the Dalitz diagrams for the electronic
density (Sec.~\ref{sec:dalitz-plots-density}).
The Dalitz diagrams suggest that the internal variables most suited for the
description of the problem are the Gronwall-Dalitz-Fabri coordinates $R,a,
\lambda$ (see (\ref{eq:a1}) of
Sec.~\ref{sec:dalitz-plots-density}) because among these coordinates the
hyperangle $\lambda$ is the ``slow variable'' as it describes the shape of the
configuration triangle.

Note that the approach employed above to explain the origin of the symmetry
(Sec.~\ref{sec:sym-origin}) is not limited to the case of planar quantum dots
for which the numerical results were presented in
Sec.~\ref{sec:numerical-results}.
Thus, one can expect that some approximate symmetries similar to that
uncovered in this article will show up in the three-dimensional case when
three electrons are confined by an arbitrary spherically symmetric potential.
Further, the consideration given in Sec.~\ref{sec:sym-origin} can be easily
generalized to the case of four- and more electrons which gives the
possibility to distinguish slow and fast variables in the corresponding wave
functions.
This, however, needs more detailed investigations.

As to the physical background of the found weak dependence of the electronic
density on the shape of the configuration triangle comparing to the dependence
on its size and area, the quantum mechanical approach does not provide any
obvious explanation.
Perhaps, the semiclassical treatment would shed some light on the physical
origin of the mentioned symmetry.

Another interesting problem would be to analyze the possible symmetry breaking
caused, for example, by the influence of an external magnetic field.
%% on the symmetry of the electronic density.
The application of the transversal magnetic field to a planar
quantum dot does not violate the circular symmetry of the Hamiltonian and,
therefore, should not change the symmetry drastically.
However, if the magnetic field has components parallel to the plane of the
quantum dot, then the symmetry of the electronic density will
be broken.
%% since the equilibrium configuration is no more an equilateral triangle.
%% Again, this is the subject of further research.
Finally, we note that effects of symmetry breaking in finite systems were
recently reviewed in \cite{rashid_prep2012}.

%%%%%%%%%%%%%%%%%%%%%%%%%%%

\section*{Acknowledgments}

This work has been supported in part by the Russian Ministry of Education and
Science under Grant No. 3.1761.2017/4.6.

%% \bibliographystyle{apsrev-titleq}
%% \bibliographystyle{unsrt}

%% \bibliography{../../mybib-hab,dots,hsh,symmetries}

\end{document}